\documentstyle[twoside,fleqn,espcrc2]{article}

\newcommand{\bea}{\begin{eqnarray}}
\newcommand{\eea}{\end{eqnarray}}
\newcommand{\beq}{\begin{equation}}
\newcommand{\eeq}{\end{equation}}
\newcommand{\nn}{\nonumber}

\newcommand{\hl}{\hline}
\newcommand{\ra}{\rightarrow}
\newcommand{\gtrsim}{\stackrel{>}{\sim}}
\newcommand{\lesssim}{\stackrel{<}{\sim}}

\title{CP and T violation in neutrino oscillations}

\author{J. Bernab\'{e}u
\address{Departamento de F\'{\i}sica Te\'{o}rica, Universidad de Valencia}
and
M.C. Ba\~{n}uls
\address{IFIC, Centro Mixto Univ. Valencia - CSIC}}

\begin{document}

\begin{abstract}
The conditions to induce appreciable CP-and T-odd effects in neutrino
oscillations are discussed. The propagation in matter leads to fake CP-and
CPT-odd asymmetries, besides a Bohm-Aharonov type modification of the
interference pattern. We study the separation of fake and genuine CP violation
by means of energy and distance dependence.
\end{abstract}

\begin{titlepage}
\onecolumn
\begin{flushright} {FTUV-00-0329}
\end{flushright}
\vskip 4cm
\centerline{\LARGE \bf
CP and T violation in neutrino oscillations\footnote{To appear in the Proceedings of TAUP 99 Workshop.}
}
\vskip 1cm
\centerline{J. Bernab\'eu
\address{Departamento de F\'{\i}sica Te\'{o}rica, Universidad de Valencia}, 
M.C. Ba\~nuls
\address{IFIC, Centro Mixto Univ. Valencia - CSIC}}
\vskip 0.5cm
\hspace*{1cm}
{$^{a \, }$ Dep. de F\'{\i}sica Te\'orica, Universidad de Valencia}\\
\hspace*{1cm}
{$^{b \, }$ IFIC, Centro Mixto Univ. Valencia - CSIC}

\vskip 1cm
\centerline{\large \bf Abstract}
\vskip 0.5 cm
The conditions to induce appreciable CP-and T-odd effects in neutrino
oscillations are discussed. The propagation in matter leads to fake CP-and
CPT-odd asymmetries, besides a Bohm-Aharonov type modification of the
interference pattern. We study the separation of fake and genuine CP violation
by means of energy and distance dependence.

\end{titlepage}
\newpage

\maketitle

Complex neutrino mixing for three neutrino families originates CP and T
violation in neutrino oscillations. 
The requirement of CPT invariance leads to the condition~\cite{Ca78}
$A\left( \overline{\alpha} \rightarrow \overline{\beta};\ t \right) 
= A^* \left( \alpha \rightarrow \beta;\ -t \right)$ for
the probability amplitude between flavour states, so that CP or T violation
effects can take place in Appearance Experiments only.

CPT invariance and the Unitarity of the Mixing Matrix imply that the
CP-odd probability
\beq
D_{\alpha \beta}=\left|  A\left(  \alpha\rightarrow\beta; \ t\right)  \right|^2
-\left| A\left ( \overline{\alpha} \rightarrow \overline{\beta};\ t\right ) \right |^2
\label{eq:cpodd}
\eeq
is unique for three flavours: $D_{e \mu}=D_{\mu \tau}=D_{\tau e}$. 
The T-odd probabilities
\bea
T_{\alpha \beta}&=&\left| A\left(\alpha \rightarrow \beta;\ t\right) \right|^2
-\left| A\left(\beta \rightarrow \alpha;\ t\right) \right|^2; \nn \\
\overline{T}_{\alpha \beta}&=&
\left|A\left( \overline{\alpha}\rightarrow \overline{\beta};\ t\right)\right|^2
-\left|A\left(\overline{\beta}\rightarrow\overline{\alpha};\ t\right)\right|^2
\label{eq:todd}
\eea
are odd functions of time~\cite{Be99} by virtue of the hermitian
character of the evolution hamiltonian. 
The last property does not apply to the effective hamiltonian of the 
$K^0 -\overline{K}^0$ system.

The oscillation terms are controlled by the phases 
$\Delta_{ij} \equiv \Delta m_{ij}^{2} L/4E$,
where $L=t$ is the distance between source and detector.
In order to generate a non-vanishing CP-odd
probability, the three families have to participate actively: in the limit
$\Delta_{12} \ll 1$, where ($1$, $2$) refer to the lowest mass eigenvalues, 
the effect tends to vanish linearly with 
$\Delta_{12}$~\footnote{We disregard the alternative solution 
where the solar neutrino oscillations are associated to an almost 
degenerate $\Delta m_{23}^2$.}. 
In addition, all mixings and the CP-phase have to be non-vanishing. 
Contrary to the CP-violating $D_{\alpha \beta}$, the CP-conserving
probabilities are not unique.
If $\Delta_{12} \ll 1$, the flavour transitions can be classified
by the mixings leading to a contribution from the main
oscillatory phase $\Delta_{23} \simeq \Delta_{13}$.
The strategy would be the selection of a 
``forbidden'' transition, i.e., an appearance channel with very low
CP-conserving probability, in order to enhance the CP-asymmetry. 
This scenario appears plausible and perhaps favoured 
by present indications on neutrino
masses and mixings from atmospheric and solar neutrino experiments. 
The use of the atmospheric results~\cite{Fu98} for 
$\left|\Delta m_{23}^2\right|$ leads to the energy dependent
\beq
\left|  \Delta_{23}\right| \simeq 4 \times 10^{-3} \left(\frac{L}{Km}\right)
\left(\frac{GeV}{E}\right)
\label{eq:de23} 
\eeq
which is of the order of unity for long-base-line experiments. 
This oscillation phase generates an ``allowed'' transition 
$\nu_{\mu} \ra \nu_{\tau}$ proportional to $s_{23}^2$
and a ``forbidden'' transition $\nu_{\mu} \ra \nu_{e}$
proportional to $s_{13}^2 s_{23}^2$
for terrestrial or atmospheric neutrinos. 
In that case, the CP-even probability is independent of $\Delta_{12}$.
The CP-odd probability is, on the contrary, 
linear in $s_{13}$ and $\Delta_{12}$. 
We conclude that the ratio $\Delta_{12}/s_{13}$ is the crucial parameter 
to induce an appreciable CP-odd asymmetry in the forbidden 
$\nu_{\mu} \rightarrow \nu_{e}$ transition.


The large-mixing-angle MSW solution~\cite{Ba98} to the solar neutrino 
data provides $\Delta m_{12}^2$ of the order of few 
$\times 10^{-5} \ eV^2$ and a mixing $\sin^2 2 \theta_{12} \gtrsim 0.7$.
For reactor neutrinos, the survival probability proceeds through
$\Delta_{13}$ (neglecting $\Delta_{12} \ll 1$) with the result
\bea
P_{\overline{\nu}_e \rightarrow \overline{\nu}_e}&=&
c_{13}^4+s_{13}^4+2 c_{13}^2 s_{13}^2 \cos(2\Delta_{13}) \nn \\
& \simeq &
1-4 s_{13}^2 \sin^2 \Delta_{23}
\eea
The CHOOZ limit~\cite{Ap98} gives $s_{13}^2 \lesssim 0.05$.
Under these circumstances, one can reach~\cite{RGH99} 
CP-odd asymmetries for the forbidden channel $\nu_{\mu} \ra \nu_{e}$ 
of the order 10-20 \%.

\begin{table}[hbt]
\setlength{\tabcolsep}{1.5pc}
\newlength{\digitwidth} \settowidth{\digitwidth}{\rm 0}
\catcode`?=\active \def?{\kern\digitwidth}
\caption{Effective mixings and oscillation lengths in matter, compared to vacuum parameters.}
\label{tab:eff}
\begin{tabular}{@{} l @{\hspace{.8cm}} c @{\hspace{.8cm}} c @{}}
\hl
 & Vacuum & Matter \\
\hl
& $\Delta_{23}$ & $\Delta_{23}$ \\
Oscillation Lengths
& $\Delta_{13}$ & $\Delta_{13}- \frac{a L}{4 E}$ \\
& $\Delta_{12}$ & $-\frac{a L}{4 E}$ \\
\hl
& $s_{23}$ & $s_{23}$\\
Mixings 
& $s_{13}$ & $\frac{s_{13}}{1-\frac{a}{a_R}}$ \\
& $s_{12}$ & $\frac{\Delta_{12}}{-\frac{a L}{4 E}} s_{12} c_{12}$ \\
\hl
\end{tabular}
\end{table}

Long-base-line experiments have large matter effects. 
They are described in the flavour basis by the effective hamiltonian
\bea
H_{\nu}\! \!& \! \! = \! \! & \! \!\frac{1}{2 E} \left \{ U 
\left (\begin{array}{@{}c@{}c@{}c@{}} 
m_1^2 & & \\ 
& m_2^2& \\
& & m_3^2
\end{array}\right )
U^+ + 
\left (\begin{array}{@{}c@{\ }c@{\ }c@{}} 
a & & \\ 
& 0& \\
& & 0
\end{array}\right )
\right \}
\nn \\
H_{\overline{\nu}} \! \!& \! \!= \! \!& \! \!\frac{1}{2 E} \left \{ U^*
\left (\begin{array}{@{}c@{}c@{}c@{}} 
m_1^2 & & \\ 
& m_2^2& \\
& & m_3^2
\end{array}\right )
U^T - 
\left (\begin{array}{@{}c@{\ }c@{\ }c@{}} 
a & & \\ 
& 0& \\
& & 0
\end{array}\right )
\right \}
\eea
where $U$ is the neutrino mixing matrix and $a$ is 
the effective potential of electron-neutrinos with electrons.
The mismatch of this charged current electron-flavour interaction 
induces a relative phase~\cite{KP87} among the electron- and the other
neutrinos which is energy independent
\beq
\frac{aL}{2E} \simeq 0.58 \times 10^{-3} \left(  \frac{L}{Km}\right); \quad
a=G \sqrt{2} N_{e} 2E
\label{eq:lbl}
\eeq
with $N_{e}$ the number density of electrons in the Earth. 
We have then the hierarchy 
$\Delta m_{23}^2 \gtrsim a \gg \Delta m_{12}^2$. 
When we diagonalize $H_{\nu}$ and $H_{\overline{\nu}}$ in 
the $\Delta_{12}=0$ limit, we observe that
matter effects break the degeneracy and
there is a resonance energy obtained from the condition 
$a=\Delta m_{23}^2 \equiv a_R$.
The effective mixing $s_{13}$ in matter is affected by the
resonance amplitude for neutrino beams. 
Although the vacuum mixing $s_{12}$ is
irrelevant in the $\Delta_{12}=0$ limit, 
the effective mixing matrix in matter
becomes determined, with $U_{e2}=0$. 
This transmutation of the vanishing
$\Delta_{12}$ in vacuum to the vanishing $U_{e2}$ in matter forbids, 
in both cases, genuine CP violating effects. 
Contrary to the ``allowed'' $\nu_{\mu} \rightarrow \nu_{\tau}$ 
transition, which is little affected by matter
effects, the forbidden transition $\nu_{\mu}\rightarrow\nu_{e}$ 
in matter has a CP-even probability
\beq
P_{\nu_{\mu}\rightarrow\nu_{e}}=4 \left( \frac{s_{13}}{1 \! - \!\frac{a}{\Delta m_{23}^2}} \right )^2 \! \! s_{23}^2
\sin^2 \left( \Delta_{23} \! - \! \frac{a L}{4E} \right)
\label{eq:pronue}
\eeq

This result shows both the enhanced probability for neutrinos (suppressed
for antineutrinos, for which $a \rightarrow -a$), 
and a modification of the interference pattern 
with an energy independent phase-shift induced by matter.
This quantum-mechanical interference provides a Bohm-Aharonov type experiment
able to detect a potential difference 
between the two ``arms'' of an interferometer. 
The interferometer is represented here by the Mixing Matrix,
the optical path difference by $\Delta_{23}$ and the potential by the
energy-independent term $\frac{a}{2E}$.

Although there are no genuine CP-violating effects in the limit 
$\Delta_{12}=0$, the medium induces fake CP-and CPT- odd effects. 
Even with fundamental CPT invariance, the survival probability 
of electron-neutrinos in matter gets modified when going to antineutrinos: 
$P_{\nu_e \ra \nu_e} \neq P_{\overline{\nu}_e \ra \overline{\nu}_e}$.
The corresponding asymmetry is, for this background effect, 
an even function of $L$. 
In order to generate genuine CP-odd effects in matter, 
one has to allow a non vanishing $\Delta_{12}$. 
The results in perturbation theory are given in Table~\ref{tab:eff},
where $\Delta_{12}$ is only maintained when needed to avoid a zero.

The CP-asymmetry in matter contains then two different terms: one fake
component induced by matter asymmetry, which is an even function of $L$, and
one genuine component, odd function of $L$, which is a true signal of CP
violation (modified by matter effects). Again the CP-odd asymmetry 
associated with the ``forbidden'' transition 
$\nu_{\mu}\rightarrow\nu_{e}$ in long-base-line
experiments is much more promising. The separation of fake and genuine
components is possible by using the energy and distance dependence 
of the asymmetry.

An alternative to the CP-asymmetry is provided by T-odd effects~\cite{AS97}. 
As matter is, in good approximation, T-symmetric, 
the T-odd asymmetry does not suffer from fake effects. 
However, its implementation will need the construction of
neutrino factories from muon-storage-rings, 
able to provide both $\nu_{\mu}$ and $\nu_{e}$ beams. 
Under the conditions discussed above, the corresponding
T-odd asymmetry for the ``forbidden'' process 
$\overline{\nu}_{\mu} \rightarrow \overline{\nu}_{e}$ is given by
\beq
A_{\not{T}} \sim -\Delta_{12} \frac{1+\frac{a}{a_R}}{s_{13}} s_{\delta}
\frac{\sin\left(  \frac{a L}{2E}\right)  }{\frac{aL}{2E}}
\label{eq:asimT}
\eeq
which can reach again appreciable values.
As in vacuum, the asymmetry varies linearly with the crucial
parameter $\Delta_{12}/s_{13}$.

If neutrinos were Majorana particles, nothing would change in the discussion
of this paper, as long as we discuss only flavour oscillations. 
The additional Majorana phases~\cite{BP83} do not enter into 
the relevant Green function 
\mbox{$<0|T\left\{\psi(x)\overline{\psi}(0)\right\}|0>$}, neither
for vacuum oscillations nor for matter~\cite{LPST87}.
One would need a Majorana propagator 
\mbox{$<0|T\left\{\psi(x) \psi^T (0)\right\}|0>$}
to be sensitive to these new ingredients.
Such a situation would affect the so-called 
``neutrino-antineutrino oscillations''.

We conclude with the comment that CP and T violation in neutrino oscillations, 
although possible in appearance experiments involving three neutrino families,
will be difficult to observe.
With a hierarchical spectrum 
to explain atmospheric and solar neutrino data, 
better prospects appear for the forbidden transition $\nu_{\mu} \ra \nu_e$
in long-base-line experiments and the large mixing angle MSW solution 
to the solar neutrino observation.
The CP-odd asymmetry becomes linear in $\Delta_{12}/s_{13}$,
the parameter associated with ``forbiddeness''.
Although matter effects break the degeneracy in $(1, \ 2)$,
the CP-odd asymmetry is still linear in $\Delta_{12}/s_{13}$
as induced by the effective mixings in matter (see Table~\ref{tab:eff}).

This work has been supported by CICYT, Spain, under Grant AEN99-0692.
One of us (M.C.B.) is indebted to the Spanish Ministry 
of Education and Culture for her fellowship.

\end{document}